\documentclass[twocolumn]{aastex6}
\usepackage{amsmath,graphics,xcolor}
\usepackage[referable]{threeparttablex}

\def\teff   {{$T_{\rm eff}$}}

\def\msun   {{M$_{\odot}$}}

\def\fk     {{$F_K$}}

\def\wsatsol{{$\omega_{\rm crit, \odot}$}}

\def\tmass  {{\it 2MASS\/}}
\def\vmk    {{$V - K_S$}}
\def\vmkz   {{$(V - K_S)_0$}}
\def\ks     {{$K_S$}}
\def\kmag   {{${K_S}$}}
\def\kepler {{\it Kepler}}
\def\ktwo   {{\it K2}}

\def\logj   {{$\log J$}}
\def\jspec  {{$j_{\rm spec}$}}
\def\logjspec  {{$\log j_{\rm spec}$}}
\def\av     {{$A_V$}}
\def\avs    {{$A_V$s}}

\begin{document}

\title{M Dwarf rotation from the \ktwo\ young clusters to the field. I. A Mass-Rotation Correlation at 10~Myr}

\email{garrett.e.somers@vanderbilt.edu}

\author{Garrett Somers\altaffilmark{1}, John Stauffer\altaffilmark{2}, Luisa Rebull\altaffilmark{2,3}, Ann Marie Cody\altaffilmark{4}, Marc Pinsonneault\altaffilmark{5,6}}
\altaffiltext{1}{Department of Physics \& Astronomy, Vanderbilt University, 6301 Stevenson Center Ln., Nashville, TN 37235, USA}
\altaffiltext{2}{Spitzer Science Center (SSC), Infrared processing and Analysis Center (IPAC), 1200 East California Blvd., California Institute of Technology, Pasadena, CA 91125, USA}
\altaffiltext{3}{Infrared Science Archive (IRSA), Infrared processing and Analysis Center (IPAC), 1200 East California Blvd., California Institute of Technology, Pasadena, CA 91125, USA}
\altaffiltext{4}{NASA Ames Research Center, Moffett Field, CA 94035, USA}
\altaffiltext{5}{Department of Astronomy, The Ohio State University, Columbus, OH 43210, USA}
\altaffiltext{6}{Center for Cosmology and Astroparticle Physics, The Ohio State University, Columbus, OH 43210, USA}

\begin{abstract}
Recent observations of the low-mass rotation distributions of the Pleiades and Praesepe clusters have revealed a ubiquitous correlation between mass and rotation, such that late M~dwarfs rotate an order-of-magnitude faster than early M~dwarfs. In this paper, we demonstrate that this mass-rotation correlation is present in the 10~Myr Upper Scorpius association, as revealed by new \ktwo\ rotation measurements. Using rotational evolution models we show that the low-mass (0.1--0.6\msun) rotation distribution of the 125~Myr Pleiades cluster can only be produced if it hosted an equally strong mass-rotation correlation at 10~Myr. This suggests that physical processes important in the early pre-main sequence (star formation, accretion, disk-locking) are primarily responsible for the M~dwarf rotation morphology, and not quirks of later angular momentum evolution. Such early mass trends must be taken into account when constructing initial conditions for future studies of stellar rotation. Finally, we show that the average M~star loses $\sim 25-40$\% of its angular momentum between 10 and 125~Myr, a figure accurately and generically predicted by modern solar-calibrated wind models. Their success rules out a lossless pre-main sequence, and validates the extrapolation of magnetic wind laws designed for solar-type stars to the low-mass regime at early times. 
\end{abstract}

\section{Introduction}\label{sec:intro}

One of the fundamental life cycles of stars is their rotational evolution. Stars spin rapidly at birth due to initial reserves of angular momentum (AM) imparted by their natal molecular clouds, and undergo dramatic rotational evolution as they contract, expand, and shed AM throughout their lifetimes \citep[e.g.][]{kraft1967,skumanich1972,barnes2003}. For decades, astronomers have been constructing physical models to understand this evolution as a function of mass and age \citep[e.g.][]{weber1967,pinsonneault1989,macgregor1991,denissenkov2010,gallet2015,somers2016}, but many uncertainties remain. The early pre-main sequence (PMS) is among the most complex and uncertain epoch of AM evolution, due to the prominent influence of accretion, the variable impact of star-disk interactions, and the observational difficulties associated with young systems \citep[e.g.][]{hartmann2001}. A final and important uncertainty is the initial AM function of protostars, as imparted by the star formation process. This function is often assumed to be weakly sensitive to mass, owing to results from higher mass stars ($M \gtrsim 0.5$\msun) in clusters such as the ONC \citep{rodriguez-ledesma2009} and h~Persei \citep{moraux2013}. 

AM studies have been propelled largely by rotation measurements in open clusters, which provide stars with a range of masses at a fixed and knowable age \citep[see][for a recent compliation]{gallet2015}, but these surveys are prone to missing very low-mass stars ($0.08-0.3$\msun), with a few notable exceptions \citep[e.g. ONC, NGC 2547, NGC 2516;][]{herbst2001,irwin2007,irwin2008b}. This situation has changed dramatically in the last few years due to the advent of long-baseline, high-cadence, space-based monitoring of rotating stars. This technique was pioneered by the {\it CoRoT} satellite \citep{baglin2006}, which produced 23~day light curves for numerous low-mass stars in the $\sim 3$~Myr old NGC~2264 cluster \citep[e.g.][]{affer2013,cody2014,stauffer2014}. {\it CoRoT} was later followed by the \kepler\ mission, which produced rotation rates for many thousands of M~dwarfs \citep{mcquillan2014}, and observed three open clusters \citep[e.g., NGC~6791, NGC~6811, and NGC~6819]{basu2011,hekker2011,corsaro2012}, in some cases detecting rotation rates of their members \citep[e.g.][]{meibom2011,meibom2015}. However, few of the field stars were young ($t < 1$~Gyr), and none this young were in the open clusters. 

Prospects for further study of early rotational evolution brightened considerably when the \kepler\ mission transitioned to \ktwo, following the failure of its second reaction wheel. The \ktwo\ mission consists of 30~minute cadence observations for a series of $\sim 78$~day pointings along the ecliptic plane, many of which contain young ($t < 1$~Gyr) open clusters and associations. These include the Pleiades, M35, Praesepe, and the Hyades. \ktwo\ has also observed at least three associations with ages $\lesssim 10$~Myr, namely Taurus-Auriga, $\rho$ Oph, and Upper Scorpius. Already, rotation periods for thousands of stars in these systems have been deduced, including $\sim 750$ members of the Pleiades \citep{rebull2016a,rebull2016b,stauffer2016}, $\sim 800$ members of Praesepe \citep{rebull2017,douglas2017}, $65$ members of Hyades \citep{douglas2017}, and $16$ brown dwarfs in the Upper Sco association \citep{scholz2015}. Moreover, our team has derived as-yet-unpublished rotation periods for hundreds of additional Upper Sco members from $\sim 0.05-2$\msun\ (Rebull et al., in prep).

One of the striking results from these studies is the strong transition to extremely rapid rotation in the low-mass regime. In both the 125~Myr Pleiades, and the $\sim 700$~Myr Praesepe and Hyades clusters, a strong mass-rotation correlation characterizes the M~dwarf distribution, with periods of 10--20~days at 0.5\msun, ranging down to 0.2--0.3~days at the sub-stellar boundary ($\sim 0.08$\msun). It is evident from studies of very young clusters that a mass-rotation correlation is present from $\sim 1$~Myr (e.g. the Orion Nebula Cluster, \citealt{herbst2001}), but it remains unclear how strongly it evolves during the PMS under the influence of magnetized winds and disk-locking. In this paper, we study this question by presenting the rotation distribution of the 10~Myr Upper Sco association. This association shows an equally strong mass-rotation trend, proving that this feature is imprinted by 10~Myr in the M~dwarf regime and does not undergo strong mass-dependent evolution thereafter due to AM loss through magnetized winds. Moreover, we show that a strong mass-rotation correlation at 10~Myr is a prerequisite for producing the trends of M~dwarf rotation in the Pleiades and Praesepe clusters. 

Our paper is organized as follows. In $\S$\ref{sec:methods}, we describe the cluster data we employ ($\S$\ref{sec:stellar_data}), how we convert these data into stellar masses ($\S$\ref{sec:masses}), our treatment of AM transport and loss ($\S$\ref{sec:models}), and our stellar models ($\S$\ref{sec:calibration}). In $\S$\ref{sec:FullRotation}, we consider first the full rotation distribution of the two clusters ($\S$\ref{sec:fgkrot}), before focusing in on the low-mass M~dwarfs, where we quantify their surface rotation ($\S$\ref{sec:mrot}) and AM ($\S$\ref{sec:AMevoltion}) evolution. In $\S$\ref{sec:EvolModels}, we forward-model the Upper Sco distribution to the age of the Pleiades, confirming both that the models are extremely reliable in this mass and age range, and that the two clusters form an evolutionary sequence. We discuss prospects for future M~dwarfs rotation studies and the 10~Myr mass-rotation correlation of Upper Sco in $\S$\ref{sec:discussion}. Finally, we summarize and conclude in $\S$\ref{sec:conclusion}.

\section{Methods}\label{sec:methods}

\begin{figure*}
\centering
\includegraphics[width=0.9\linewidth]{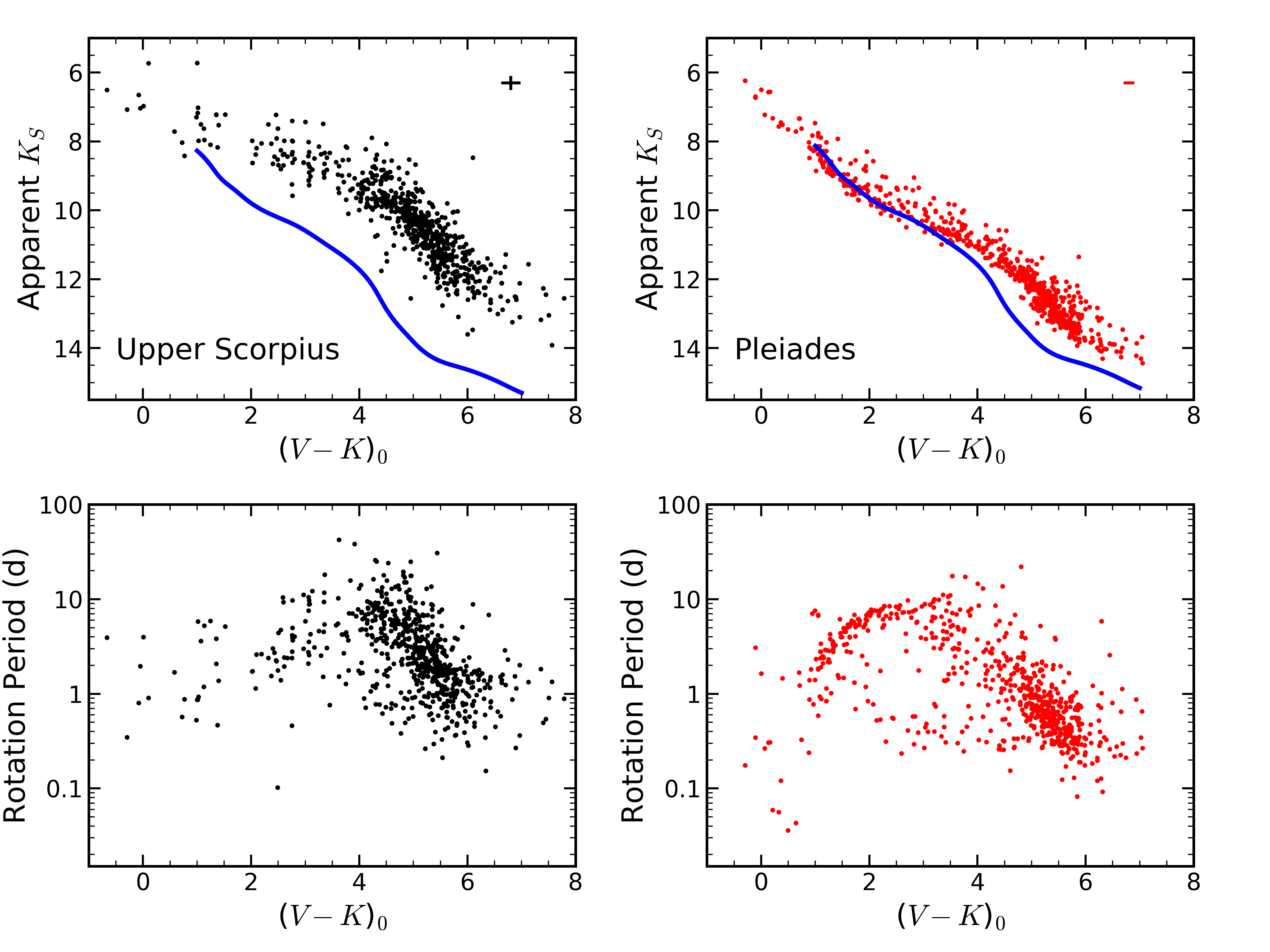}
\caption{
Color-magnitude (top) and rotation period (bottom) diagrams for Upper Sco (left) and the Pleiades (right). These \kmag\ and \vmkz\ colors include the reddening corrections described in $\S$\ref{sec:stellar_data}-\ref{sec:masses}. The blue line represents the zero-age main sequence from \citet{baraffe2015} isochrones, demonstrating that the Pleiades is older than Upper Sco. Typical errors, resulting from uncertainties on the photometry, reddening corrections, and distances, are shown in the top right.
\label{fig:VmkPlots}
}
\end{figure*}

\subsection{Stellar and cluster data}\label{sec:stellar_data}

\noindent

We begin by describing the cluster parameters used in this paper. We adopt for the Pleiades a distance of 136~pc \citep{melis2014}, an age of 125~Myr \citep{stauffer1998}, and a V-band extinction of \av\ = 0.12 \citep[e.g.][]{an2007} with ${\rm R}_V = 3.1$ and wavelength dependence given by \citet{cardelli1989}. For Upper Sco, we take a distance of 145~pc \citep{deZeeuw1999}, and estimate \av\ values individually for each star ($\S$\ref{sec:masses}). The age of Upper Sco has been a contentious subject---early studies found ages of around 5~Myr based on comparisons of stellar models and their HR-diagram locations \citep{preibisch2002,slesnick2008}, but more recent work  with updated stellar parameters for F-stars determined a median age of $\sim 11$~Myr \citep{pecaut2012}. These discrepancies reflect discordant HR-diagram ages for the lower and high-mass populations in the cluster \citep{herczeg2015}. One potential resolution to this discrepancy is that magnetic activity and starspots have altered the fundamental parameters of lower mass stars \citep[e.g.][]{feiden2013,somers2015a,macdonald2017,somers2017}, making them appear younger in the HR diagram, an noted in other young clusters \citep{jackson2014,jackson2016,jeffries2017}. When considering these effects, \citet{feiden2016} found a consensus age of 10~Myr for the cluster---we adopt this age.

From \citet{rebull2016a}, we collect Pleiades rotation periods, \tmass\ \kmag\ measurements, and estimated \vmk\ values. Notably, these \vmk\ values are an assortment of real V-band measurements compared to {\it 2MASS\/} \ks, and transformations from other color bands (mainly from SDSS $g-$\ks\ and $r-$\ks). A detailed description of the Upper Sco data will be presented in a forthcoming paper (Rebull et al., in prep), but the analysis is quite similar to that of the \ktwo\ Pleiades rotation rates \citep[see][]{stauffer2017}. These data are plotted in Fig. \ref{fig:VmkPlots}, with the reddening corrections described above. The blue line shows an approximate zero-age main sequence (ZAMS) from the models of \citet{baraffe2015}, demonstrating the relative evolutionary states of the two clusters. It is clear that the M~dwarfs (\vmk\ $\gtrsim 3.9$) are far closer to the ZAMS in the Pleiades than in Upper Sco. Upper Sco members of 0.1\msun\ have $\sim 3.4 \times$ their ZAMS radius, whereas Pleiads are only $\sim 1.4 \times$ larger. At 0.4 and 0.6\msun, Pleiads are within 3\% of their ZAMS radii, but Upper Sco members are still $2.1$ and $1.7\times$ larger. 

In this paper, we are interested predominantly in the evolution of single stars, and will discuss the evolution of binary rotation rates in an upcoming paper (Stauffer et al., in prep). Accordingly, we adopt only those stars with a single strong peak in their periodogram---multiple peaks for M~dwarfs are interpreted as possible signatures of binarity \citep{rebull2016b}. Additional photometric, spectroscopic, and visual binaries identified in the Pleiades, as aggregated by \citet{rebull2016a}, have also been excluded.

\begin{figure*}
\centering
\includegraphics[width=0.9\linewidth]{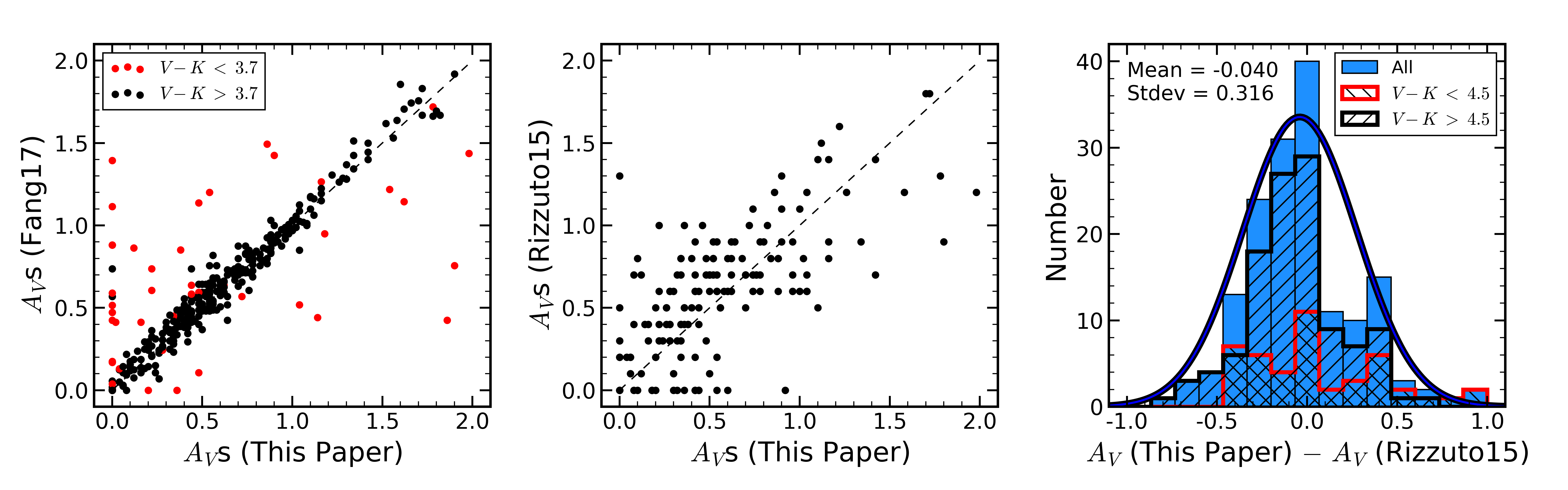}
\caption{
{\it Left:} Comparison between \avs\ derive in this work and \avs\ from \citet{fang2017}. Stars redder than $V-K = 3.7$ ($\sim$0.8\msun) agree well, but bluer stars do not. {\it Center:} Comparison between our \avs\ and those from \citet{rizzuto2015}. The values correlate with one another, with a spread of 0.1-0.3~mag around the one-to-one line. {\it Right:} Histogram of the difference between our \avs\ and those from \citet{rizzuto2015}. The Gaussian mean and $\sigma$ for all stars (blue) are given in the top left. The 0.32~mag spread of the Gaussian reflects the quadrature sum of our errors and the 0.2~mag errors from \citet{rizzuto2015}, suggesting an uncertainty of $\sim 0.25$~mag for our derived \avs. Black and red histograms show the hotter and cooler distributions respectively, demonstrating that the most deviant outliers are indeed hot stars.
\label{fig:CompAVs}
}
\end{figure*}

\subsection{Mass estimates}\label{sec:masses}

In order to compare the clusters to our rotating models we elect to work in mass coordinates, necessitating the color-mass conversions described here. For the Pleiades, we first plotted the cluster in the color-magnitude diagram (CMD) of \vmkz\ vs. $M_{K_S}$. Next, we drew by eye a line at the approximate location of the single-star main sequence. For each star in the cluster, we projected it down (or up) in the CMD onto the single star sequence, to find the $M_{K_S}$ corresponding to its \vmkz\ value. Finally, we interpolated in the 120~Myr isochrones of \citet{baraffe2015} to find the mass corresponding to this value of $M_{K_S}$. This procedure is preferable to a simple \vmkz--mass mapping, as the \citet{baraffe2015} isochrones do not exactly match the locus of Pleiades stars. Moreover, by projecting the stars down to the single-star sequence, we lessen the impact of undetected low-mass binaries on the final mass estimate. 

Deriving masses for Upper Sco presents an additional complication. Due to its young age, substantial knots of dust from the natal molecular cloud remain in the vicinity of Upper Sco, producing significant star-to-star differential extinction within the cluster. We therefore estimate line-of-sight $A_V$ values for each individual star, using the method of \citet{fang2017}. We first tabulate \tmass\ $J$, $H$, and \ks\ magnitudes for each star, and combine these values with the \vmk\ estimates discussed above to derive estimated $V-J$ and $V-H$ colors. Next, we assume the relative reddenings given in each photometric band \citet{fiorucci2003},

\[
A_V = \frac{E(V-K_S)}{0.89} = \frac{E(V-H)}{0.83} = \frac{E(V-J)}{0.73}.
\]

\noindent
With these constraints, we search for the value of \av\ which brings the de-reddened colors of each individual star into best agreement with PMS colors from \citet{pecaut2013}. Stars with derived V-band extinctions greater than 2.0 are discarded for simplicity, and \av\ = 0 is set as a lower bound. Once each star is de-reddened, we follow the mass-derivation steps outlined for the Pleiades, using the 10~Myr isochrones from \citet{baraffe2015}.  

As a check of our extinction measurements, we compare our derived \avs\ to the values obtained by \citet[][private communication]{fang2017} in the left panel of Fig. \ref{fig:CompAVs}. We find that for lower mass stars (\vmk$> 3.7$, $\sim 0.8$\msun\ at 10~Myr), the agreement is very good, showing that we have successfully replicated their method in the mass range of interest for this paper. For higher mass stars, there appears to be no correlation between the two values. This could be because we have used a color-\teff\ relation, whereas \citet{fang2017} used a spectral type-\teff\ relation, leading to inconsistencies when TiO bands become too weak to aid spectral typing.

We also compare in the center panel our \avs\ to those from \citet{rizzuto2015}, who matched extinguished spectral templates to optical spectra of Upper Sco members. It is evident that the values correlate with one another, but there exists a spread of a few tenths of a mag around the one-to-one line. \citet{rizzuto2015} notes typical \av\ errors of 0.2~mag, so we show in the right panel a histogram (blue) of the difference between our and their \avs\ for each star present in both samples.
Fitting a Gaussian to the histogram gives a mean $\Delta$\av\ of -0.04~mag, and a standard deviation of 0.32~mag. The distribution is formally inconsistent with a Gaussian per an Anderson-Darling test, but the results can still give guidance in estimating uncertainties on our \avs. The standard deviation of this histogram reflects the quadrature sum of errors from our \avs\ and errors from the \avs\ of \citet{rizzuto2015}, and thus Fig. \ref{fig:CompAVs} implies that the errors on our values are $\sim 0.25$~mag, not substantially larger than the comparison sample. The black and red histograms compare the cooler and hotter portions, showing a somewhat larger offset among the hottest stars. A fit to the black histogram gives errors of just 0.2~mag on our \avs\ for the cooler stars. We conclude that our \av\ errors are likely comparable to those of \citet{rizzuto2015}.

These \av\ errors, along with distance uncertainties and noise in the photometry, formally propagate to $\sim 0.1$\msun\ for M stars in our isochrone-based mass calculation. For a complimentary estimate of the uncertainties, we derived masses using our method for the primaries in the Upper Sco eclipsing binaries discussed in \citet{rizzuto2016}. When comparing to their mass measurements, we find an average difference of $-0.06 \pm 0.11$\msun, in good agreement within the uncertainties derived from error propagation. 

\subsection{Angular momentum transport and loss}\label{sec:models}

To determine whether modern stellar wind laws predict the evolution of M~dwarf rotation rates from the age of Upper Sco to the age of the Pleiades, we test several AM evolution models which vary in their treatment of magnetized winds and internal AM transport. Our wind laws take the form given by Eq. \ref{eqn:windlaw},

\begin{equation} 
 \displaystyle \frac{dJ}{dt} = - \left\{
	\begin{array}{l l}
	 \displaystyle F_K K_M \left(\frac{\omega \omega_{crit}^2}{\omega_{\odot}^3}\right), \quad \omega_{crit} < \omega \\
	 \displaystyle F_K K_M \left(\frac{\omega^3}{\omega_{\odot}^3}\right), \quad \omega_{crit} \geq \omega \\
	\end{array} \right. 
	\label{eqn:windlaw}
\end{equation}

Here, $J$ is stellar AM, $\omega$ is the rotation rate, $\omega_{\odot}$ is the solar rotation rate (taken as $2.86 \times 10^{-6}\ {\rm s}^{-1}$), $F_K$ is the overall normalization of the wind law, and $\omega_{crit}$ reflects the saturation threshold, a rotation rate below which stars lose AM as the cube of their rotation, and above which stars lose AM linearly with rotation \citep{krishnamurthi1997}. The precise value of the saturation threshold scales with the depth of the convection zone for stars of different masses ($\omega_{crit} = \omega_{crit,\odot} \frac{\tau_{\rm CZ,\odot}}{\tau_{\rm CZ,*}}$; \citealt{sills2000}), thus enforcing a fixed saturation Rossby number\footnote{The Rossby number is equal to the ratio of the rotation period to the convective overturn timescale, an empirically important scaling for the generation of magnetic fields and surface magnetic activity.} in accordance with chromospheric and coronal data \citep[e.g.][]{noyes1984,pizzolato2003,wright2011}. \fk\ and \wsatsol\ are not {\it a priori} known, and so must be calibrated for each individual wind law ($\S$\ref{sec:calibration}).

The final parameter, $K_M$ [erg], is a product of different structural variables of the star, and is determined by the evolutionary models we use. The precise combination of structure variables entering into $K_M$ for each wind law differs, thus producing different predictions. In this paper, we test three different wind laws from the literature. First, we consider the \citet{krishnamurthi1997} formulation of the classic magnetic wind law advanced by \citet{kawaler1988}. This model gives

\begin{equation}
K_M = 2.6 \times 10^{30}
\left(\frac{R}{R_{\odot}}\right)^{\frac{1}{2}} 
\left(\frac{M}{M_{\odot}}\right)^{-\frac{1}{2}},
\end{equation}

\noindent
where $M$ and $R$ are the mass and radius of the star. Second, we test the physical formulation of magnetic mass loss promoted by \citet{matt2012}, also modified to include a Rossby-scaled saturation threshold \citep{vansaders2013}. This gives,

\begin{equation}
  \begin{split}
 \displaystyle K_M = 
 1.3 \times 10^{30}
 \left(\frac{R}{R_{\odot}}\right)^{3.1} 
 \left(\frac{M}{M_{\odot}}\right)^{-0.22} 
 \left(\frac{L}{L_{\odot}}\right)^{0.56} \\
 \times \left(\frac{P_{phot}}{P_{phot,\odot}}\right)^{0.44} 
 \left(\frac{\tau_{\rm CZ}}{\tau_{\rm CZ,\odot}}\right)^{2}
 c(\omega),
 \end{split}
\end{equation}

\noindent
where $L$ is the stellar luminosity, $P_{phot}$ is the gas pressure at the photosphere, $\tau_{\rm CZ}$ is the convective overturn timescale, and $c(\omega)$ is a centrifugal correction term\footnote{While \citet{vansaders2013} set this value = 1 for simplicity, we have adopted the form given in Eq. 8 of \citet{matt2012}}. Finally, we test the empirically-calibrated wind law of \citet{matt2015}, which gives,

\begin{equation}
\displaystyle K_M =
9.5 \times 10^{30}
\left(\frac{R}{R_{\odot}}\right)^{3.1}
\left(\frac{M}{M_{\odot}}\right)^{0.5}
\left(\frac{\tau_{\rm CZ}}{\tau_{\rm CZ,\odot}}\right)^{2}.
\end{equation}

\noindent
The physical interpretations of each wind law will not be discussed here, and we refer the interested reader to the source papers.

A second important component of AM evolution models is the transport of AM between the interior layers of stars. The speed at which loss at the surface is communicated to the deeper layers has strong consequences for the time-dependent evolution of the observable surface rotation rate. The impact of internal AM transport on the morphological features of stellar surface rotation as a function of age has been explored in detail in many previous studies \citep[e.g.][and many more]{pinsonneault1989,krishnamurthi1997,denissenkov2010,gallet2013,somers2016}, and we refer the reader to those papers, and references within. 

We consider in this paper two limiting cases of internal AM transport. The first is solid body rotation, and the second is a core-envelope re-coupling framework, treated in the two-zone approximation \citep{macgregor1991}, with a variable coupling timescale $\tau_{CE}$ as determined by \citet{lanzafame2015}. We note that these authors only measured $\tau_{CE}$ down to $\sim 0.6$\msun, so we fit an exponential to their values and extrapolate down to the fully convective boundary. This extrapolation produces a very long re-coupling timescale ($>$1~Gyr), and so can be thought of as an extreme limiting case of core-envelope de-coupling, in contrast to the alternate extreme limiting case of solid body rotation. These differences predominantly matter for stars above the fully convective boundary of $\sim 0.35$\msun, because lower mass stars are treated as solid bodies at all times. As a result, the details of internal transport have little direct impact on the rotation rates of low-mass M~dwarfs, and only factor into our results inasmuch as they influence the wind law calibration ($\S$\ref{sec:calibration}). 

\begin{figure*}
\centering
\includegraphics[width=0.9\linewidth]{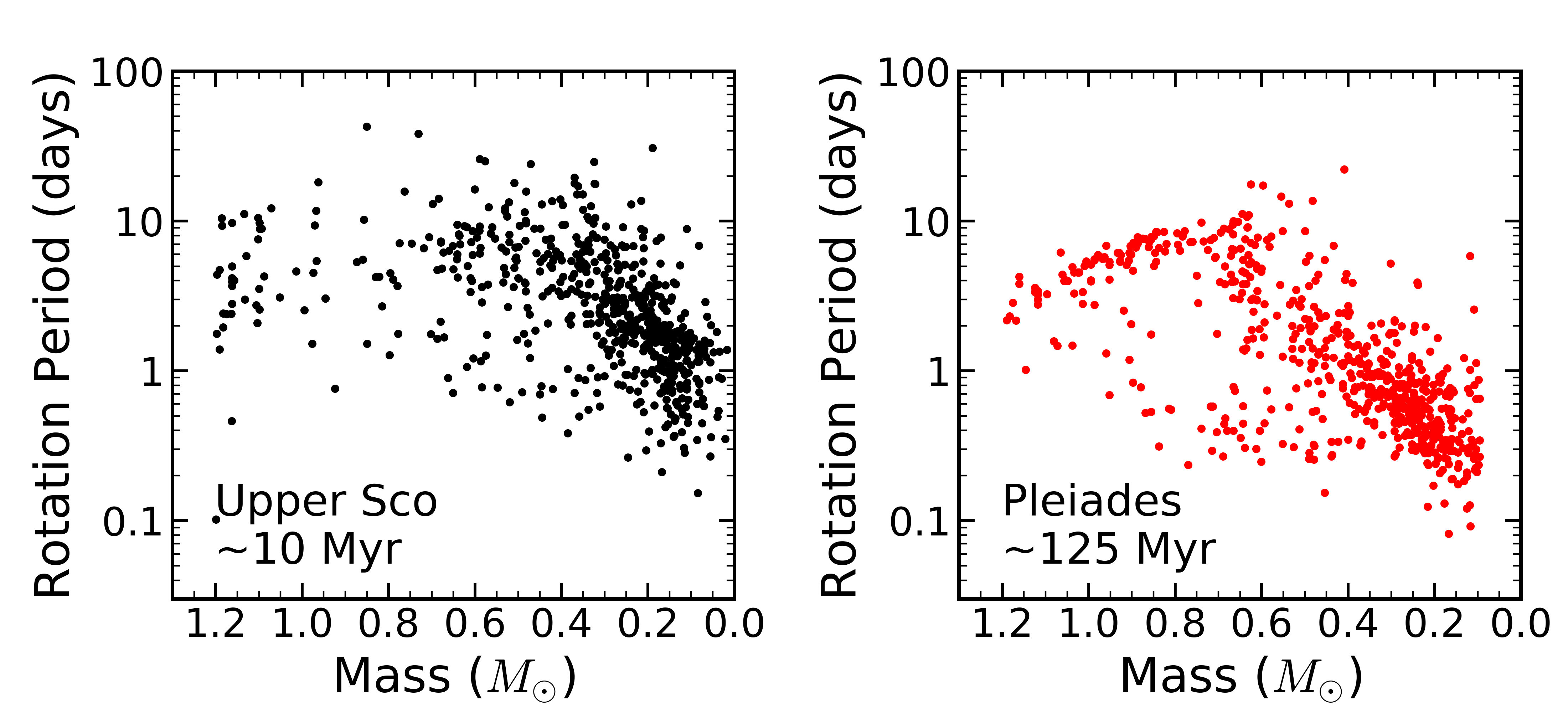}
\caption{Derived masses and rotation rates for 0.05--1.2\msun, from Upper Sco (left) and the Pleiades (right). Both clusters show a strong correlation between mass and rotation rate in the M~dwarf regime.
\label{fig:FGKMs}
}
\end{figure*}

\begin{table}[b]
	\caption{AM model calibration}
	\label{tab:calib}
	\begin{centering}
	\begin{threeparttable}
	    \begin{tabular}{lcll} 
		    \hline
		    \hline
	    	Wind Law & Transport\tnote{a} & $F_K$ & $\omega_{\rm crit,\odot}$  \\
		    \hline
		    Kawaler88 & SB & 2.73 & 15.8$\omega_{\odot}$ \\
		    Matt12    & SB & 9.0 & 14.7$\omega_{\odot}$ \\
		    Matt15    & SB & 0.75 & 10.1$\omega_{\odot}$ \\
		    Kawaler88 & RC & 8.0 & 8.0$\omega_{\odot}$ \\
		    Matt12    & RC & 14.0 & 12.2$\omega_{\odot}$ \\
		    Matt15    & RC & 1.8 & 8.0$\omega_{\odot}$ \\
		    \hline
	    \end{tabular}
	    \begin{tablenotes}
		\item[a] SB = solid body; RC = Core-envelope recoupling. See $\S$\ref{sec:models} for details.
	    \end{tablenotes}
    \end{threeparttable}
    \end{centering}
\end{table}

Finally, we note that AM loss models have historically been constructed to reproduce the rotation rates of solar-mass stars, and not M~dwarfs. However, their inherent scalings based on mass, radius, luminosity, and other stellar properties can be computed for M~dwarfs, and thus constitute pure predictions for this mass regime. For this reason, we elect to calibrate our models on solar-mass stars, a process we describe in the next section.

\subsection{Stellar models}\label{sec:calibration}

To create our stellar models, we first generate non-rotating evolutionary tracks for mass between 0.05 and 1.2\msun, using the Yale Rotating Evolution Code ({\tt YREC}). We adopt the physics and composition described in \citet{somers2014}, assuming a present day solar photospheric mixture given by \citet{grevesse1998}. The helium abundance and mixing length parameter are calibrated to reproduce the solar radius and luminosity at 4.57~Gyr for a solar mass model. We then use the rotation code {\tt rotevol}, which takes as input non-rotating stellar tracks, a starting rotation period, and an AM loss law, and determines the angular momentum evolution thereafter \citep[see Sec. 2.1 in][]{vansaders2013}.

Our adopted wind laws have two free parameters which must be calibrated against empirical data. These are the overall normalization of the AM loss (\fk), and the rotation rate at which the Sun enters the saturated regime (\wsatsol). To perform the calibration, we follow the procedure laid out in \citet{somers2015b}. This method involves tuning \wsatsol\ such that stars with rapid initial rotation rates reproduce the fastest rotating 1\msun\ stars in a young ($t \sim 100$~Myr) cluster, and tuning \fk\ so that stars with average initial rotation rates produce the rotation rate of the slowest stars in an older cluster. For this purpose, we use the stars of mass 0.95--1.05\msun\ in the \ktwo\ rotation distributions of the Pleiades and Praesepe \citep{rebull2017}, and determine fast and median initial conditions from the rotation rates given by \citet{moraux2013} for the 13~Myr open cluster h~Persei \citep[see][for more details]{somers2015b}. The final calibrated parameters are listed in Table \ref{tab:calib}.

Finally, we briefly discuss our method for computing the convective overturn timescale $\tau_{\rm CZ}$. In our methodology, $\tau_{\rm CZ}$ is derived directly from our models. There are two traditional approaches to this: either an integrated value throughout the convection zone, or a local value above base of the convection zone (see discussion in \citealt{kim1996}). In the usual local formulation, one evaluates the pressure scale height at the convection zone base ($H_{\rm CZ}$), evaluates the convective velocity exactly this distance above the convection zone base ($V_{\rm CZ+H}$), and defines $\tau_{\rm CZ} = H_{\rm CZ}/V_{\rm CZ+H}$. This formulation creates problems for fully convective stars, where the pressure scale height diverges at the center. Instead, we adopt a variant in {\tt YREC}: we evaluate the pressure scale height through the convection zone, and search for the location where the base of the surface convection zone (or the center of the star in the case of a fully-convective object) is one pressure scale high below. This behaves similarly to the traditional local method for thin surface convection zones but is properly defined and more stable for fully convective stars.

\begin{figure*}
\centering
\includegraphics[width=0.9\linewidth]{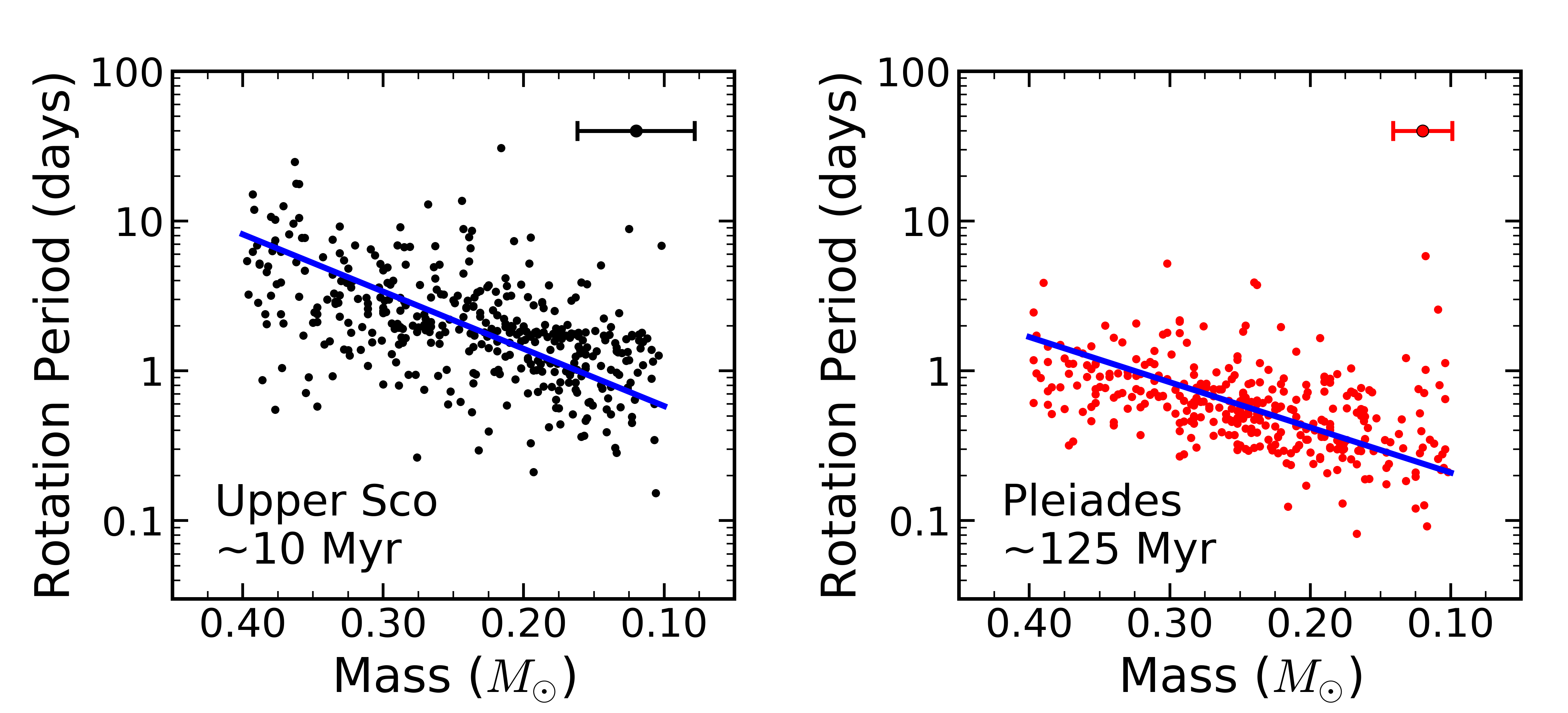}
\caption{The same as Fig. \ref{fig:FGKMs}, but zoomed into the stars between 0.1--0.4\msun. This represents the approximate range of the stark mass-rotation correlation among M~dwarfs in the younger cluster. Blue lines represent a best fit exponential relation between the two quantities (see text). The Pleiades relation is shallower, perhaps reflecting the more rapid contraction of higher mass stars on the PMS. Typical error bars are shown in the top right, reflecting propagated mass errors from uncertainties in the reddening and photometry. 
\label{fig:JustMs}
}
\end{figure*}

\section{The First 120 Million Years}\label{sec:FullRotation}

We now present our derived mass-rotation relations for the two clusters, and compare them to one another. First, we discuss the bulk properties of the entire rotation data sets, before focusing in on just the M~dwarf regime and discussing the mass-rotation correlation. We then calculate the AM content of the cluster M~dwarfs, and derive the total AM lost between 10 and 125~Myr. These values will serve as benchmarks for our forward models, discussed in the next section.

\subsection{Empirical Rotation Of FGK Stars}\label{sec:fgkrot}

Fig. \ref{fig:FGKMs} shows rotation rates for the FGKM stars in Upper Sco and the Pleiades, plotted against the masses derived in $\S$\ref{sec:masses}. Considering first Upper Sco, we see that the rotation pattern can be divided into two regimes. For stars more massive than 0.4\msun, Upper Sco hosts a fairly featureless and flat distribution of rotation rates between $\sim 0.3-30$~days. This appears in line with rotation rates down to the K~dwarf regime for younger clusters, such as the Orion Nebula Cluster \citep[e.g.][]{herbst2001} and NGC 2264 \citep[e.g.][]{lamm2004}, though there seems to be little evidence of the nascent slow and rapid branches appearing in some older clusters \citep[e.g. h Persei $\sim 13$~Myr][]{moraux2013}. This could reflect the relatively low number of stars in the higher mass bins of our sample, though other recent studies of Upper Sco have not found a bimodality either \citep{mellon2017}.

By Pleiades age, we begin to see the emergence of the well-known slow rotator sequence, resulting from the gradual convergence of FGK~star rotation rates during spin down \citep[e.g.][]{epstein2014}. This feature ranges from $\sim 2$~days at 1.2\msun\ to $\sim 10$~days at 0.6\msun. These reflect descendents of the slower rotating stars in Upper Sco. By contrast, little convergence has occurred for stars between 0.6--0.4\msun, and the Pleiades shows a dispersion of similar magnitude as that in Upper Sco. This is a consequence of the young age of the cluster; a dominant converged sequence emerges over the next few hundreds of Myrs in this mass range (e.g. M37, \citealt{hartman2009}; Praesepe, \citealt{rebull2017}).

\begin{figure*}
\centering
\includegraphics[width=0.9\linewidth]{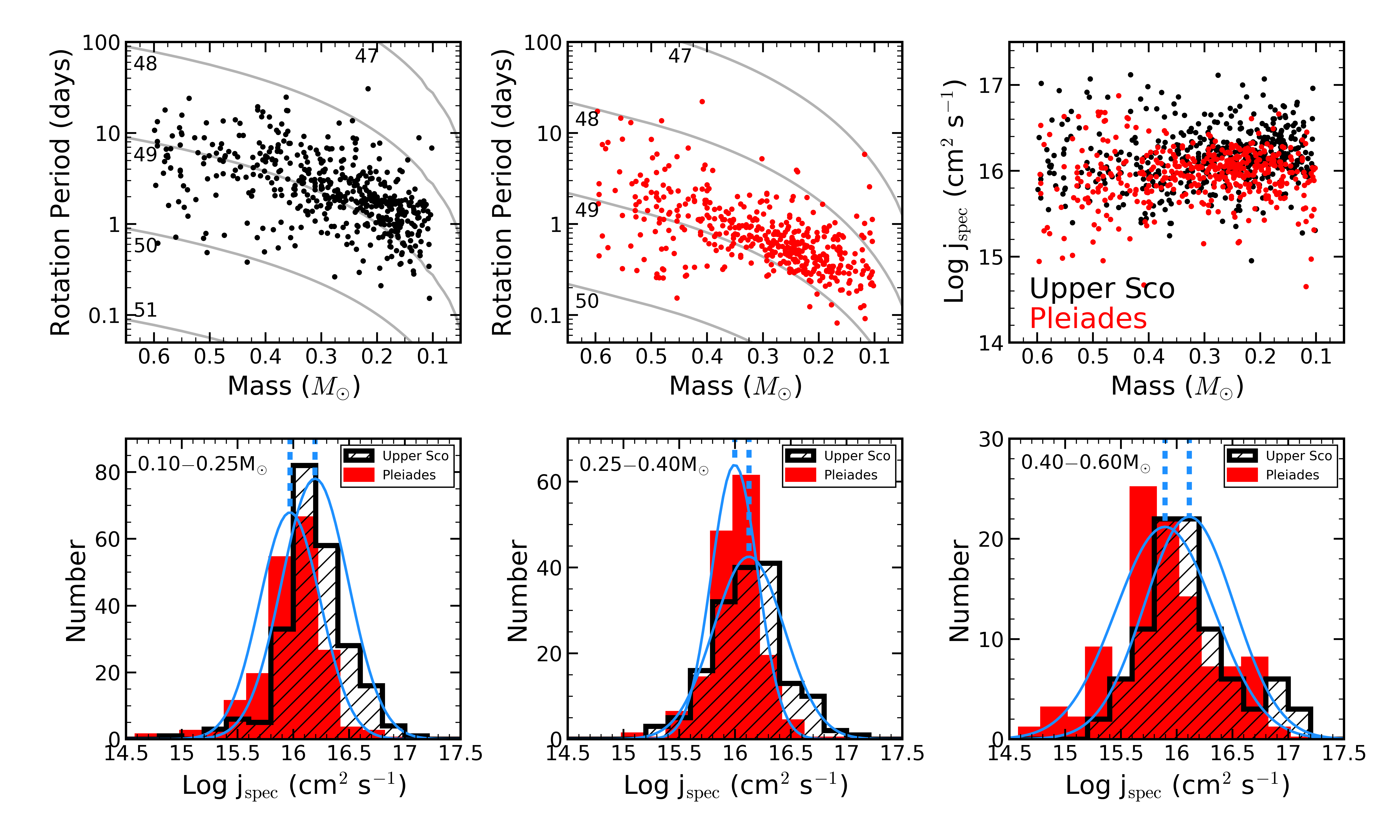}
\caption{An empirical look at the angular momentum content of Upper Sco and the Pleiades. {\it Top left and center:} Rotation rates of the two clusters, plotted against lines of constant angular momentum from our stellar models (grey), assuming solid body rotation. Numbers reflect the \logj\ values of each line. {\it Top right:} \logjspec\ values for each star in the two clusters, determined by interpolation between the grey lines in the left column. The strong mass-rotation correlation corresponds to an essentially flat \jspec\ function with mass. {\it Bottom:} Histograms of \logjspec\ for Upper Sco (black) and Pleiades (red), in three different mass bins. The blue curves are Gaussian fits to each histogram, demonstrating that in each bin, the average Pleiades member has $\sim 25-40$\% less specific AM than the average Upper Sco member.
\label{fig:UscoPleiadesAM}
}
\end{figure*}

Overall, the FGK pattern of Upper Sco falls in line with previous observations, and shows no surprises relative to expectations, with the possible exception of a missing bimodality in Upper Sco. Next, we turn to the M star rotation distribution.

\subsection{Empirical Rotation Of M Stars}\label{sec:mrot}

Below 0.4\msun, we find a prominent relationship between mass and rotation among the M~dwarfs of Upper Sco, in contrast to the flat pattern characterizing the high mass end. The sign of the trend is such that 0.1\msun\ stars rotate significantly faster on average than 0.4\msun\ stars, with an apparent exponential relationship. While the vast majority lie within $\sim 3\times$ this mean trend, there exist a handful of slowly-rotating stars around 0.1--0.2\msun, and a handful of rapidly-rotating stars around 0.4\msun, filling in the two quadrants bisected by the exponential relationship. These stars could indicate that the general range of rotation rates seen at F-, G-, and K- type also persists in the M~dwarf regime, and the mass-rotation trend is some additional component lying on top. However, contamination by binaries, inaccurate extinction corrections, and non-member contaminants of order 5--10\% of the sample likely contribute to this scatter as well.

The M~dwarfs of the Pleiades exhibit a similar morphology. Below $\sim 0.4$\msun, we again find an exponential increase in rotation rate towards lower masses, with a handful of stars populating the other two quadrants. The predominant change in the rotation distribution from 10 to 125~Myr is not morphological as seen for the FGK stars, but instead is a generic increase in the average rotation rate. This shift is clearly a consequence of PMS contraction inducing spin up due to AM conservation. 

To bring these rotation distributions into sharper relief, we zoom into the stars between 0.1-0.4\msun, the approximate range of the strong Upper Sco mass-rotation trend, in Fig. \ref{fig:JustMs}. In this zoom plot, the similarity of the Upper Sco and Pleiades distributions is even more evident. To numerically vet the similarity, we fit each cluster with an exponential of form $\log_{10}(P_{\rm rot}) = a \times M + b$, using an ordinary least-squares bisector method to determine $a$ and $b$. We find $\log_{10}(P_{\rm rot}) = 3.82 M - 0.62$ for Upper Sco, and $\log_{10}(P_{\rm rot}) = 3.02 M - 0.98$ for the Pleiades. The slightly shallower Pleiades slope may result from the faster Hayashi contraction, and thus more rapid spin-up, of the higher mass stars. Both correlations are highly significant, per a Kendall Tau test (p = $3 \times 10^{-32}$ and $10^{-25}$). Nonetheless, the precise values of the slopes should be taken with caution as the overall normalization of the Upper Sco mass scale remains uncertain.

The logarithmic standard deviation around the mean trend in Upper Sco is $\sim 0.30$~dex, larger than the $\sim 0.24$~dex dispersion in the Pleiades. The tighter Pleiades distribution likely indicates superior reddening corrections, mass estimates, and membership, but could also result from the greater fractional age spread in the younger cluster. It is also likely that our Upper Sco data set contains more binaries, which could bias mass estimates of individual objects. 

\subsection{Angular Momentum Loss}\label{sec:AMevoltion}

To more directly measure AM loss during the PMS, we next determine the AM content of each star in our sample. To do this, we extract the moment of inertia $I$ at 10 and 125~Myr as a function of mass from the stellar evolution models discussed in $\S$\ref{sec:calibration}. Assuming solid-body rotation\footnote{This assumption implies that the resulting value is a lower limit on the total AM of the object, because for partially-convective stars the core may in fact be more rapidly rotating than the surface.}, we can calculate angular momentum $J$ [g~cm$^2$~s$^{-1}$] by multiplying $I$ by the angular velocity of each star, related to the rotation period by $\omega_{*} = 2 \pi / P_{\rm rot}$. To illustrate, we plot contours of fixed \logj\ alongside the M~dwarf rotation distributions of the two clusters in the top left and center of Fig. \ref{fig:UscoPleiadesAM}. In both clusters, the slope of the contours are similar to the mass-rotation trends up to 0.6\msun. Stars in both clusters congregate around \logj$ = 48-49.5$, showing remarkable similarity in the average AM content. Above 0.4\msun, the strong mass trend seems to vanish, and a relatively flat distribution takes hold, ranging between \logj$ = 48-50$. 

We next converted these values to specific AM (\jspec) by interpolating between these contours to the mass and rotation rate of each star, and dividing the resulting $J$ of each star by its mass. The resulting \logjspec\ values are plotted versus mass in the top right. In these coordinates, the mass trend of the mean value is almost completely absent. This is confirmed by Kendall Tau tests, which show a much reduced significance for a correlation in the Upper Sco stars (p = $3 \times 10^{-5}$), and no significant correlation in the Pleiades stars (p = 0.06). It thus appears that the M~dwarfs of these two clusters are essentially flat in specific AM (see $\S$\ref{sec:discussion}). The lack of significant mass-dependence for both clusters suggests that whatever AM evolution occurs between 10 and 125~Myr is largely insensitive to mass. This fact implies that stars in this mass range remain magnetically saturated up to the age of the Pleiades, producing self-similarity in the AM distributions \citep{tinker2002}.

Finally, in the bottom row of Fig. \ref{fig:UscoPleiadesAM} we plot histograms of the \logjspec\ values in three different mass bins. In each plot, we find that the Upper Sco peaks are statistically significantly higher in \logjspec\ by $\sim 0.12-0.20$~dex. This signifies the small amount of AM loss occurring due to stellar winds on the PMS. To estimate the average and dispersion, we fit Gaussian distributions to all six peaks, shown as blue lines in the bottom row of Fig. \ref{fig:UscoPleiadesAM}. The Gaussian centers and standard deviations for Upper Sco and the Pleiades are $16.20 \pm 0.29$ and $15.97 \pm 0.27$ for the low mass bin, $16.12 \pm 0.32$ and $16.00 \pm 0.21$ for the middle bin, and $16.11 \pm 0.40$ and $15.89 \pm 0.44$ for the high mass bin. This equates to AM losses of approximately 40\%, 25\%, and 39\% from the lowest to highest mass bin.

These histograms demonstrate that stars must lose AM between 10 and 125~Myr, ruling out lossless models of the post-disk PMS. We will return to this issue in the next section when considering the predictions of our models.

\begin{figure*}
\centering
\includegraphics[width=0.9\linewidth]{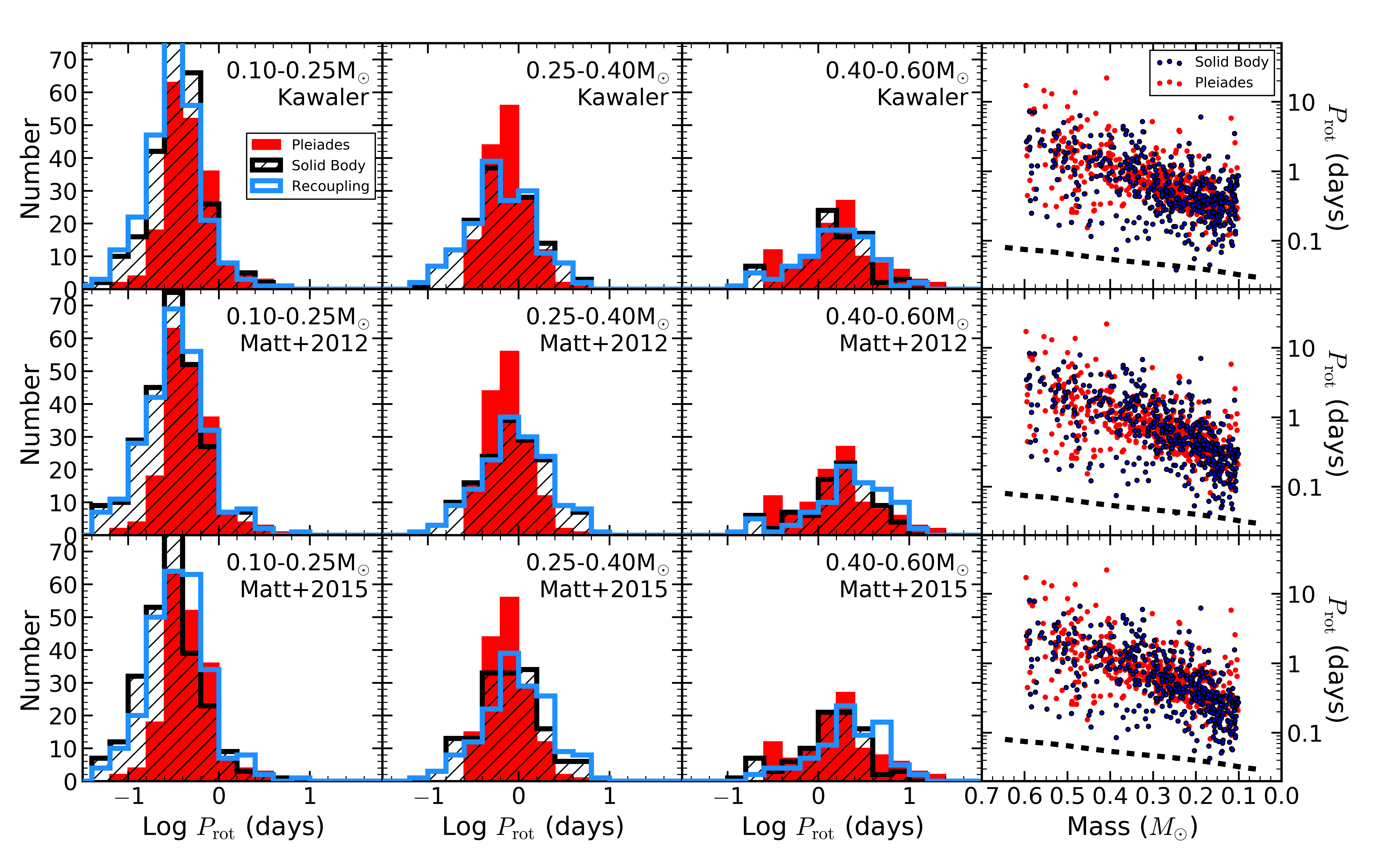}
\caption{Results from forward-modelling the Upper Sco rotation distribution to the age of the Pleiades. In each panel of the first three columns, the red histogram reflects the Pleiades rotation distribution, the black histogram is a solid body forward model, and the blue histogram is a core-envelope recoupling forward model (see $\S$\ref{sec:models}). Each row uses a different wind law: \citet{kawaler1988} on the top, \citet{matt2012} in the middle, and \citet{matt2015} on the bottom. The agreement between the peak and spread of the forward models is in general good, suggesting the wind laws are successfully predicting early M~dwarf rotational evolution. The right column compares the Pleiades rotation distribution in red to the solid body forward models of each wind law in blue. Black dashed line show the approximate break up velocity at 125~Myr, and represents an approximate lower envelope of rotation at all masses.
\label{fig:FwdModels}
}
\end{figure*}

\section{Evolutionary Models}\label{sec:EvolModels}

Having described the salient details of the Upper Sco and Pleiades M~dwarf patterns, we now turn to the predictions of rotating evolutionary models.  Previous work has tested models of AM evolution against the higher mass ($M \gtrsim 0.5$\msun) stars in both Upper Sco \citep[e.g.][]{mellon2017,ansdell2017} and the Pleiades \citep[e.g.][]{stauffer1984,soderblom1993c,bouvier1997,sills2000,coker2016} so we consider our investigation the lower-mass analog to these results.

Here we address two main questions. 1) Do our wind laws predict the $\sim 3 \times$ increase in the average rotation rate, and the $\sim 25-40$\% decrease in the average AM, empirically detected from 10 to 125 Myr? 2) How necessary is the observed mass-rotation correlation in Upper Sco to producing the observed Pleiades pattern? As we will show, evolutionary models generically predict the features of the Pleiades rotation distribution with great accuracy -- with the exception of the older \citet{kawaler1988} wind law at the low mass end -- but require a mass-rotation pattern to be fixed by the age of Upper Sco in order to be compatible.

\begin{figure}
\centering
\includegraphics[width=0.9\linewidth]{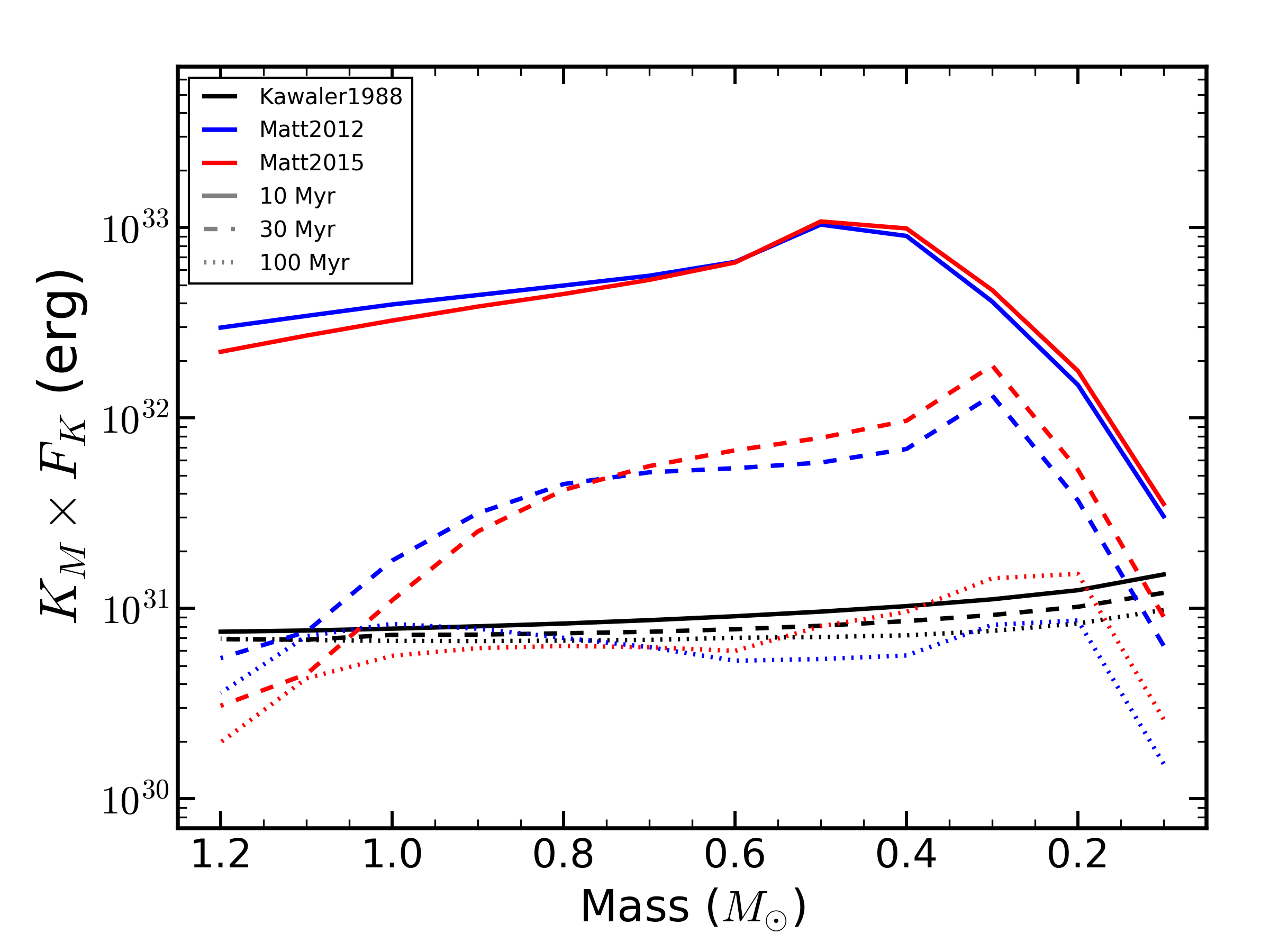}
\caption{Calibrated scaling constants (see $\S$\ref{sec:models}) in the three wind laws at 10, 30, and 100 Myr (solid, dashed, dotted). The two \citeauthor{matt2012} wind laws are far stronger before 100~Myr, but the brief duration of this early epoch and the extremely large moments of inertia of PMS stars means that the early rotational evolution of stars do not differ greatly in the different treatments.
\label{fig:kmvals}
}
\end{figure}

\begin{figure*}
\centering
\includegraphics[width=0.9\linewidth]{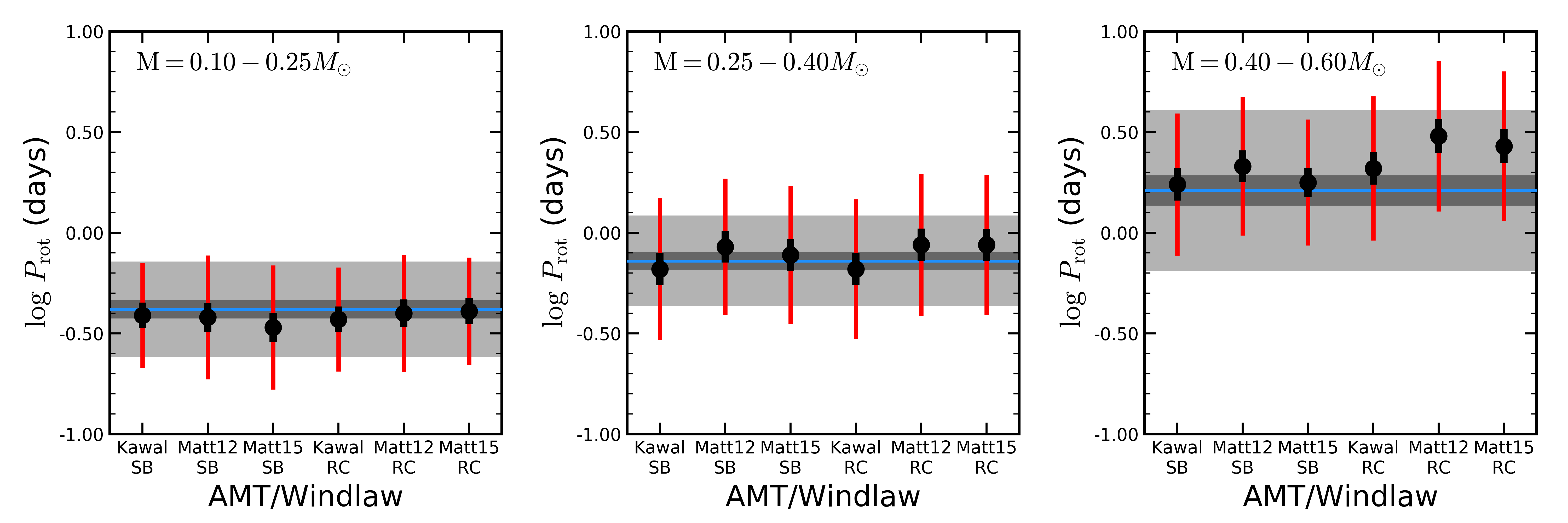}
\caption{Graphical representation of the goodness-of-fit of the forward modeled Upper Sco distribution for each wind law in Fig. \ref{fig:FwdModels}. The blue line represents the mean rotation rate, with the dark and light shaded regions reflecting the standard error of the mean and standard deviation for Gaussian fits to the Pleiades histograms in Fig. \ref{fig:FwdModels}. In each of the three mass bins, we compare solid body (SB) and core-envelope recoupling (RC) models, whose average rotation rates are shown as black points, and whose standard error of the mean and deviation are shown as black and red error bars. It is clear that in most cases, the average of the models match the average Pleiades rotation rate within the standard errors of the mean, suggesting the forward models reproduce early AM evolution well. The dispersions match well in the lowest mass bin, are too large in the middle bin, and are too small in the final bin (see the text for discussion).
\label{fig:GaussianResults}
}
\end{figure*}

\subsection{Forward Modeling}\label{sec:FwdModeling}

Using the machinery and calibrations discussed in $\S$\ref{sec:methods}, we compute forward models of the rotation rates of Upper Sco members of $0.1 M_{\odot} \leq M \leq 0.6$\msun, begun at 10 Myr and ending at 125 Myr. These models account for the spin up of stars due to PMS contraction and AM conservation, the loss of AM at the surface due to winds, and the transport of AM in the stellar interiors.  In Fig. \ref{fig:FwdModels}, we show the resulting forward models for solid body (black) and core-envelope recoupling (blue) models with the \citet{kawaler1988}, \citet{matt2012}, and \citet{matt2015} wind laws. These are compared as histograms to the Pleiades cluster data (red) in three different mass bins. In the right-most column, we plot individual stars in the mass-rotation plane, comparing the Pleiades cluster data in red to the solid body forward model in black (the recoupling forward models look very similar, and for simplicity are not shown). The dashed black line corresponds to the break up rotation period for Pleiades-age stars, calculated with the equation given in \citet{bouvier2013}, $V_{break\ up} = \sqrt{\frac{2}{3} \frac{G M_*}{R_*}}$. Overall, these figures demonstrate the following.

First, we find excellent agreement between the average rotation rate of the forward model predictions and the average rotation rate in each mass bin. The peak rotation periods match extremely well, deviating only in the higher mass bin for the re-coupling versions of the two \citeauthor{matt2012} wind laws. Consequently, the mass-rotation slope shown in the right panel is accurately predicted along the full M~dwarf sequence for every wind law. Moreover, the forward-modeled Upper Sco stars seem to approach, but not violate, the break up velocity represented by the dashed line. 

Second, the predicted dispersions about the mean in the forward models differ in accuracy from bin to bin. In the lowest mass bin, the predicted dispersions are somewhat larger in the forward model, due to the higher number of rapid rotators.  In the middle mass bin, the model clearly over-predicts the dispersion for the Pleiades. In the final mass bin, the dispersions are similar. These features can be traced to differences between the cluster rotation distributions as discussed in $\S$\ref{sec:mrot}. 

Third, we find striking similarity between the predictions of the various wind laws during this early epoch. What discrepancies we see are primarily due to minor differences in the calibration of the model parameters, which were fit to reproduce Pleiades and Praesepe rotation trends at 1\msun, and thus reflect details of the rotational evolution of very different stars. This similarity may seem surprising given the differences in the scaling parameters employed by each wind law, particularly the inclusion of the convective overturn timescale in the two \citeauthor{matt2012} formulations ($\S$\ref{sec:models}). Indeed, the strength of AM loss differs quite a lot at early times between the different wind laws (Fig. \ref{fig:kmvals}). However, PMS stars have such large moments of inertia that the resulting rotational changes are quite small relative to those induced by contraction. Moreover, this early epoch of enhanced AM loss in the \citeauthor{matt2012} wind laws is so brief that the total lost AM is not substantial. For these reasons, early predictions from different wind laws cannot differ from one another substantially once calibration has occurred.

These impressions are supported numerically by Fig. \ref{fig:GaussianResults}, which compares the average Pleiades period in each mass bin (blue line), and its standard error of the mean and deviation (shaded regions), to the average rotation periods, standard errors of the mean, and standard deviations for each forward model. In all three mass bins, the models generally agree within $\lesssim 1.5 \sigma$ of the Pleiades distribution, with the poorest fits being only slightly worse ($\sim 2 \sigma$). The dispersion about the mean matches reasonably well in the lowest bin, but are clearly larger in the second mass bin as previously noted. Notably, the dispersions from the core-envelope re-coupling models are larger than those of the solid body models in the highest mass bin. The re-coupling model dispersions match better the empirical spread, perhaps indicating the burgeoning importance of core-envelope de-coupling above the full-convective boundary.

\begin{figure*}
\centering
\includegraphics[width=0.9\linewidth]{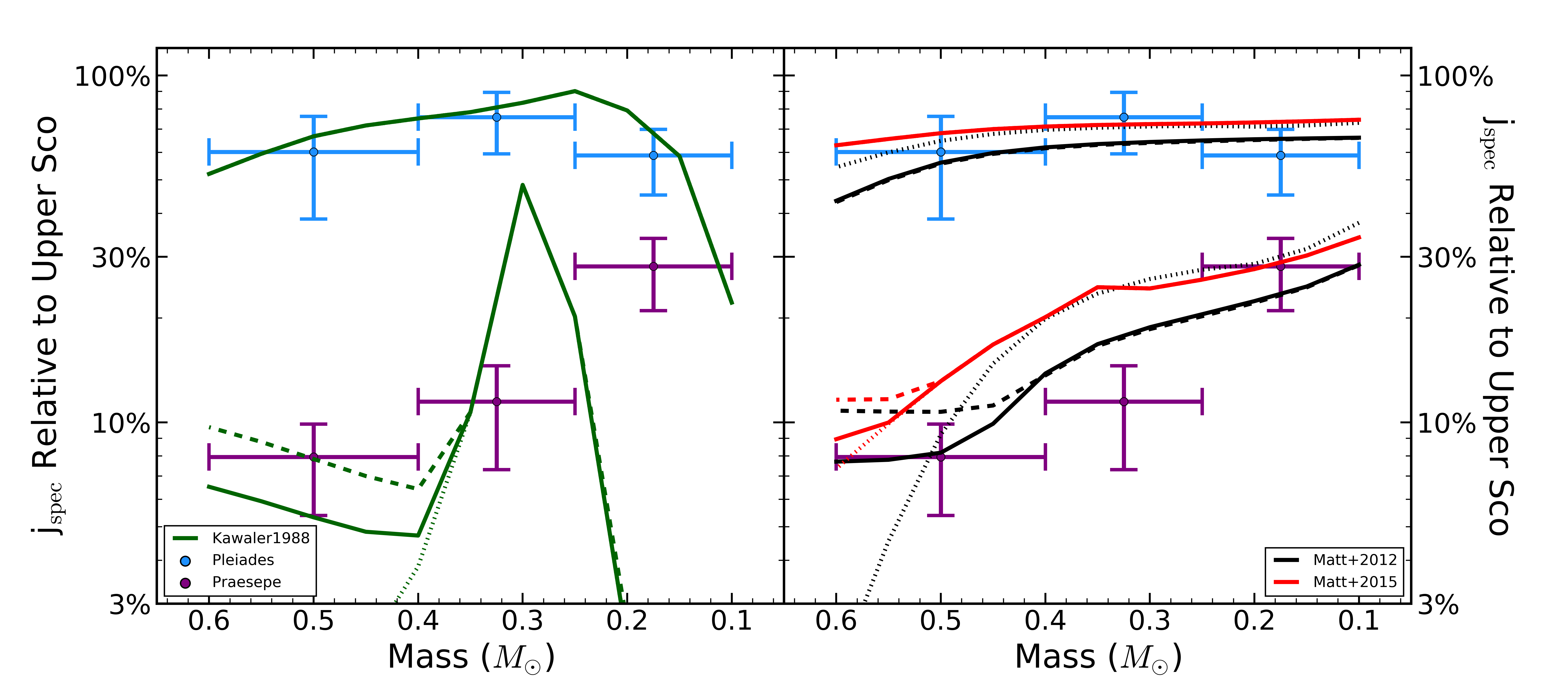}
\caption{A comparison between the angular momentum content of the Pleiades (blue) and Praesepe (purple), relative to Upper Sco, in three mass bins. Vertical error bars represents the standard deviation of AM values in each mass bin. On the left, these are compared to \citet{kawaler1988} solid body forward models, for fast (dotted), median (solid), and slow (dashed) initial conditions (see $\S$\ref{sec:methods}). The right is the same, except using the \citeauthor{matt2012} wind law. The \citeauthor{kawaler1988} models perform poorly at the low-mass end, but the \citeauthor{matt2012} models predict reasonably well the AM decrement for each data point, except perhaps the median Praesepe bin (see text).
\label{fig:AMmodels}
}
\end{figure*}

Our forward models for the \citet{kawaler1988} wind law successfully predict the Pleiades distribution down to $\sim 0.15$\msun, but break down at the very bottom of our distribution. To demonstrate this, we show in Fig. \ref{fig:AMmodels} empirical estimates of the amount of AM lost from Pleiades stars since the age of Upper Sco (blue), assuming solid body rotation. We also show a similar calculation for stars in the $\sim$700~Myr Praesepe cluster (purple), with rotation rates from \citet{rebull2017}, and masses estimated in the same fashion as the Pleiades. On the left, we compare these data to predictions from \citet{kawaler1988} solid-body models (green) for fast, median, and slow initial rotation rates.\footnote{These limits correspond to 10th, 50th, and 90th percentile rotation rates between 0.4--0.6\msun\ in h~Persei \citep{moraux2013}.}. For the two higher mass bins these predictions are quite accurate at 125~Myr, but they predict too rapid spin down at the very low mass end. These problems become even more pronounced at the age of Praesepe, where the anticipated spin down is much stronger than observed in the data at the low mass end, and far less than predicted in the middle mass bin.

On the right, we compare the cluster averages to predictions from \citet{matt2012} and \citet{matt2015} solid-body models. These models predict much better the evolving AM distribution for both clusters, likely due to their more complex dependencies on the stellar properties ($\S$\ref{sec:models}). The one exception is the middle mass bin in Praesepe, which is significantly more drained of AM than the models predict, suggesting that stars around 0.4\msun\ spin down faster than expected after the age of the Pleiades -- this has been noted by \citet{douglas2017}, and will be explored in detail in an upcoming paper (Somers et al., in prep). Although our forward modeling exercises above found little difference in the predictions of the models, a direct look at the evolving AM budget provides strong justification for using the newer models of AM loss.

\subsection{Other Initial Conditions}\label{sec:otherstars} 

\begin{figure*}
\centering
\includegraphics[width=0.9\linewidth]{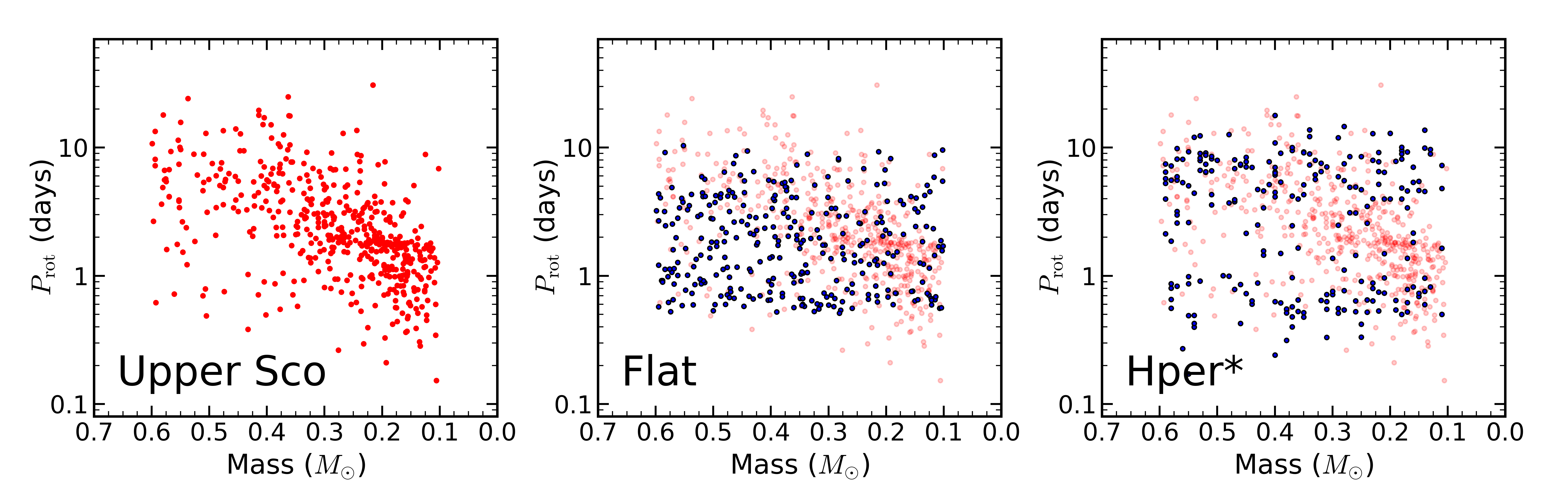}
\caption{Alternative possible initial conditions for AM evolution, compared to the Upper Sco distribution. {\it Left}: The Upper Sco distribution from this work. {\it Center}: A mass-independent, log flat rotation distribution at 10~Myr in blue, compared to Upper Sco in light red. {\it Right:} A rotation distribution mimicking the h~Persei distribution (see text) in blue, compared to Upper Sco in light red.
\label{fig:AlternateStartInit}
}
\end{figure*}

The excellent match of the mass-rotation slopes in the M~dwarf regime raises a question: can the Pleiades pattern be predicted with a flatter mass-rotation relationship, the sort expressed by higher mass stars? If not, then this supports the reliability of our mass and rotation measurements for Upper Sco. To test this notion, we computed forward models with two alternative sets of initial conditions, shown in Fig. \ref{fig:AlternateStartInit}.

1) A log-flat initial rotation distribution independent of mass. Starting conditions were generated by randomly drawing mass between 0.1-0.6\msun, and rotation rates at 10~Myr from 0.6--10~days, following \citet{matt2015}.
    
2) An initial distribution mimicking the rotation distribution observed for higher mass stars in h~Persei \citep{moraux2013}. This distribution displays a clear bimodal behavior, with fast and slow rotating branches. The observed rotation pattern extends down to just $\sim 0.4$\msun, but is relatively flat with respect to mass, so we simply subtracted 0.3\msun\ from the derived masses of \citet{moraux2013} to create a synthetic h~Persei distribution for the desired mass range.\footnote{While this fake distribution is emphatically not compatible with our derived Upper Sco rotation rates, it represents a reasonable expectation based on the h~Persei observations.}

\begin{figure*}
\centering
\includegraphics[width=0.9\linewidth]{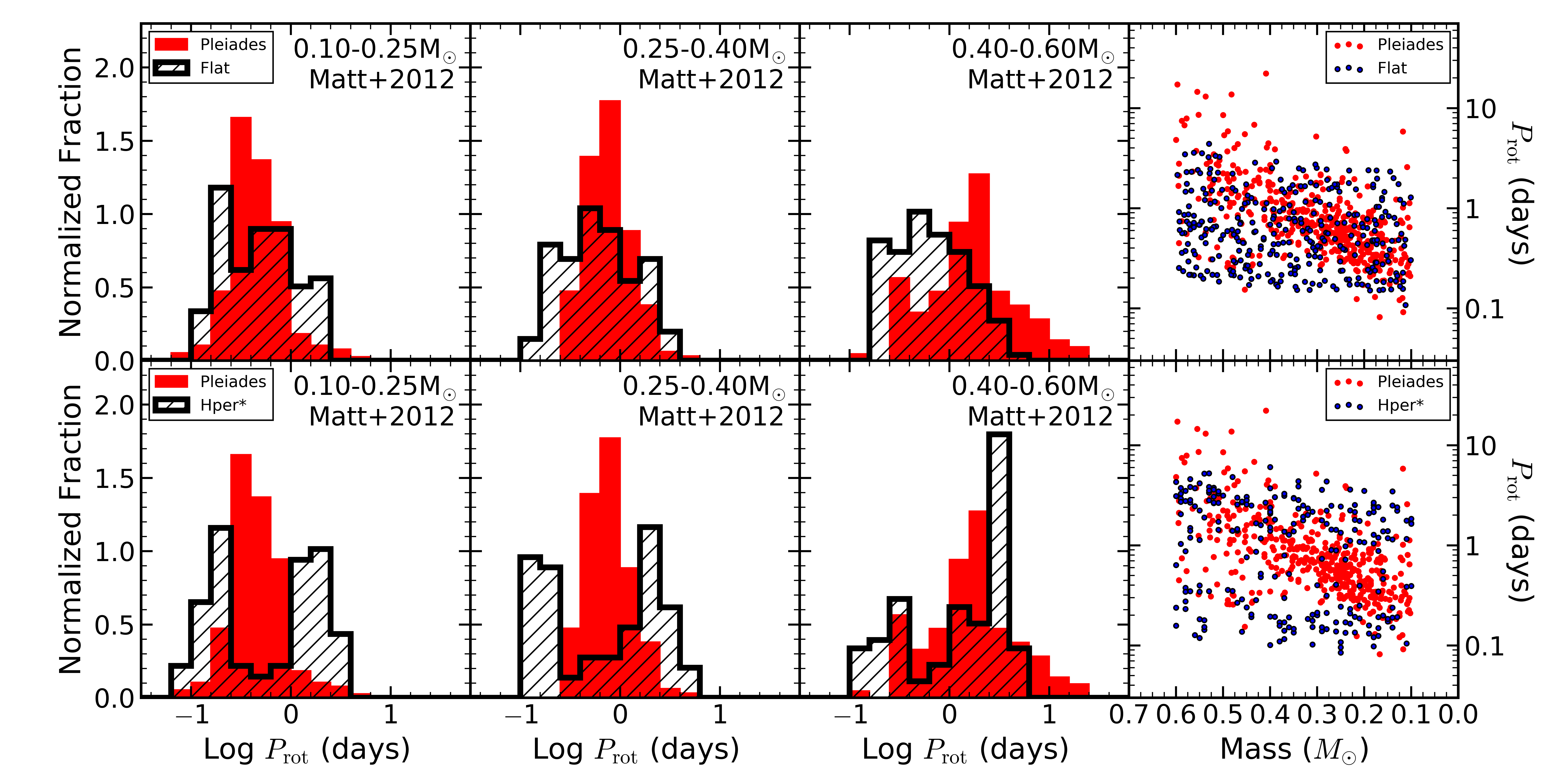}
\caption{Similar to Fig. \ref{fig:FwdModels}, but with different rotation initial conditions. {\it Top row:} Comparing the Pleiades rotation distribution (red) in three mass bins to a forward model initialized at 10~Myr with a mass-independent, log flat rotation distribution. This model produces dispersions that are too large, and does not produce a strong mass-rotation trend. {\it Bottom row:} Comparing the Pleiades to a forward model initialized at 10~Myr with a rotation distribution mimicking the h~Persei distribution (see text). This distribution produces a double peaked feature, strongly inconsistent with the uni-modal nature of the Pleiades M~dwarfs. 
\label{fig:AlternateStarts}
}
\end{figure*}

We compare these forward models to the Pleiades in Fig. \ref{fig:AlternateStarts}. In the top row, the flat distribution shows a viable central value for the two lower mass bins, but a far flatter and less peaked overall distribution. This is a consequence of the log-flat nature of the initial conditions, which conflicts with the highly peaked and mass-dependent center of Pleiades. In the third mass bin, we find far too many rapidly rotating stars in the forward model. This is because the slower stars are not as prominent in this synthetic initial distribution as they are in Upper Sco.

The bottom row compares the mock h~Per distribution to the Pleiades. The bimodality of the initial conditions is preserved at 125~Myr, strongly contradicting the observed Pleiades distribution, which remains quite peaked at this age. As for the highest mass bin, we find too few slow rotators and too many fast rotators. However, the models again predict a double-peaked structure, which we see hints of in the Pleiades distribution, albeit with low significance. This would not be surprising, as 0.4--0.6\msun\ is well within the mass range where bimodality is observed in young clusters. This model does, however, predict far more rapid rotators than are found in the Pleiades \citep[see][]{coker2016}.

We conclude that the M~dwarf structure in the Pleiades does not arise naturally from flat initial conditions as a consequence of AM evolution after 10~Myr, and instead requires a strong mass-rotation correlation to be imprinted on the early PMS. Once set in, AM evolutionary models generically predict the general shape of the Pleiades distribution.

\begin{figure*}
\begin{centering}
\includegraphics[width=0.85\linewidth]{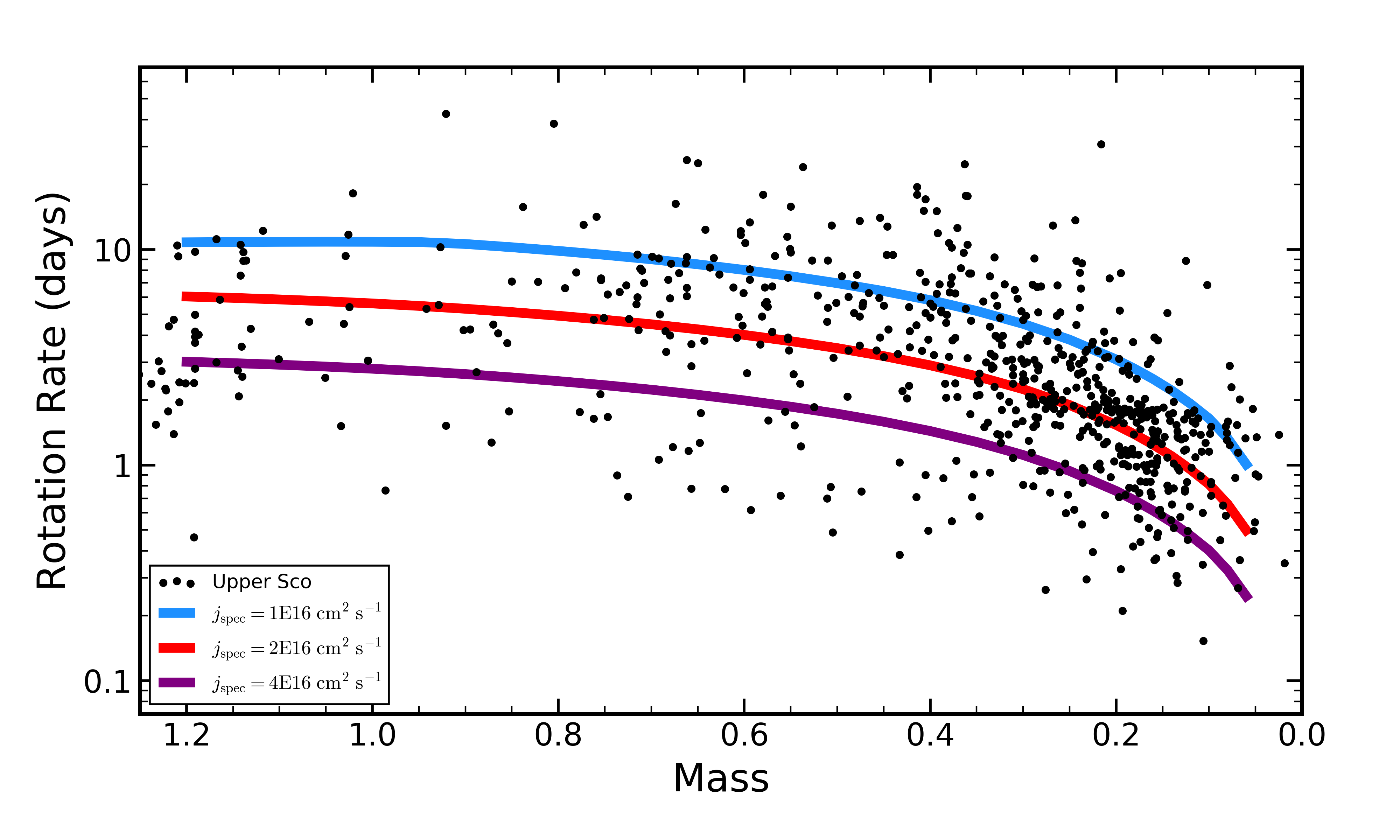}
\caption{A comparison between the Upper Sco rotation distribution, and solid-body \citet{matt2012} models initialized with different values of fixed specific angular momentum, given by the key. Mass-insensitive \jspec\ initial conditions are motivated by the observed distribution of rotation rates in the ONC \citep{herbst2001}, and predict a strong mass-rotation trend at low masses, but a much weaker trend near 1\msun.
\label{fig:OncUpperSco}
}
\end{centering}
\end{figure*}

\section{Discussion}\label{sec:discussion}

Prior to 2016, only a few dozen stars near the sub-stellar boundary with known ages above 100~Myr had detected rotation rates \citep[e.g.][]{irwin2011}. The \ktwo\ mission has provided the first real trove of such stars, and with it a unique opportunity for understanding the physics governing their spin down. We have already shown the potential of these new observations, demonstrating that modern magnetized wind laws make accurate predictions for the evolving morphology of the PMS M~dwarf AM distribution, outperforming older treatments based on more simplistic physics. However, the fidelity of these models at early ages is largely due to the simplistic physics of PMS contraction, and does not suggest that the models make accurate and precise predictions thereafter, when the details of the wind laws and internal AM transport become all important. In upcoming work (Somers et al., in prep), we will examine the longer term spin down of low-mass stars using Praesepe rotation rates from \ktwo\ \citep{rebull2017} and field rotation rates from M~dwarfs in the solar neighborhood \citep[e.g.][]{stelzer2016,newton2016a}.

The predominant feature of the Upper Sco rotation distribution is the mass-rotation correlation below 0.4\msun. This trend has been hinted at previously by \citet{dahm2012} using $v \sin i$s, and by \citet{scholz2015} for brown dwarfs, but have been conclusively demonstrated in this paper. It is also evident in other young clusters such as the Orion Nebula Cluster \citep{herbst2001}, NGC~2264 \citep{lamm2005}, NGC~2362 \citep{irwin2008a}, and $\sigma$~Ori \citep{cody2010}, among others. Despite its ubiquity, its origin and evolution remains unclear, and appears contradictory at times. \citet{henderson2012} suggested that the slope of the correlation evolves with age, and their predictions match the slope we infer for Upper Sco ($\S$\ref{sec:mrot}), but they found little sign of the correlation in the very young ($\sim 1$~Myr) NGC~6360. However, the feature is already strong in the equally young ($1-2$~Myr) ONC \citep{herbst2001}, suggesting that it must be present from birth in at least some clusters. Could the Upper Sco trend have evolved with age between 1 and 10~Myr, or was it imprinted at birth?

A hint to the resolution of this question may come from the ONC. \citet{herbst2001} noted that in this cluster, the median specific angular momentum (\jspec) of stars between $0.1-1$\msun\ is only weakly dependent on mass. For this mass range, they found \jspec\ $= 1-5 \times 10^{16}\ {\rm cm^2\ s^{-1}}$, corresponding on the high mass end to the slow-rotating, ``disk-locked'' branch. A constant \jspec\ implies that absolute $J$ scales linearly with mass. However, the moment of inertia $I$ scales with $M R^2$, suggesting that at fixed \jspec, $\omega = J/I$ actually increases towards lower mass. Thus, the mass-rotation correlation at 10~Myr could arise directly from the mass-insensitivity of \jspec\ at $1-2$~Myr. Similar arguments were explored by \citet{stauffer2016} who compared the Pleiades distribution to NGC~2264, finding similar mass-rotation slopes in the M~dwarf regime, and that about half of the total AM of low-mass stars is lost due to disk-locking and magnetic winds between 3 and 125~Myr. Our rotating models allow us to build upon these tests, and extend the analysis to the rich and high quality Upper Sco data set.

To test this idea, we calculate three sets of solid body, \citet{matt2012} evolutionary models from $0.1-1.2$\msun\ (see $\S$\ref{sec:methods}), initiated at 1~Myr with a fixed \jspec\ values of 1, 2, and 4 $\times 10^{16}\ {\rm cm^2\ s^{-1}}$. We refer to these here as our low, median, and high $J$ initial conditions. The models at 10~Myr are plotted against the Upper Sco rotation distribution in Fig. \ref{fig:OncUpperSco}. It is clear that in this scenario the expected rotation rate increases rather sharply below about 0.4\msun, with the median-$J$ model neatly bisecting the M~dwarf mass-rotation correlation, and the low and high-$J$ models bracketing the densest portion of the low mass distribution. At higher masses, the trend flattens substantially, and centers around 5-10 days for 1\msun, demonstrating that the fastest rotating low-mass stars have similar \jspec\ values as the stars which underwent substantial disc-locking in their early lifetimes.

It is evident that a steep mass-rotation relation would develop for initial conditions with a weakly mass-dependent \jspec. However, it is far beyond the scope of this paper to test the important effects of disc-locking, accretion, and non-zero age spreads, which must also influence the 10~Myr rotation morphology. We suggest instead that, given the natural progression from the mass-rotation relation in the ONC \citep{herbst2001} to the present day Upper Sco, this feature likely develops quite early, and retains much of its \jspec\ structure throughout the disc-locking phase. We hope that the stark demonstration of this feature in Upper Sco will be of aid to future studies on the nature of the star formation process.

\section{Summary and Conclusions}\label{sec:conclusion}

With the advent of long-baseline, high-cadence, space-based observations of low mass stars by mission such as {\it CoRoT} and \ktwo, measurements of large sets of stars in nearby open clusters down to the sub-stellar boundary have become numerous. These data present a golden opportunity for testing our models of angular momentum evolution for low-mass M~dwarfs. To this end, we have examined the \ktwo\ M~dwarf rotation distributions in Upper Scorpius and the Pleiades, seeking to understand both the generic features of open cluster rotation patterns at 10~Myr, and whether modern stellar evolution models accurately predict the evolution of angular momentum during the PMS.

First, we found that Upper Sco hosts a prominent correlation between mass and rotation rate below $\sim 0.4$\msun, in the sense that the average rotation rate increases towards lower mass. Although structure in this mass range appears quite common in young clusters, the feature in Upper Sco may be the strongest example of the phenomenon discovered to date due to the high quality of the \ktwo\ data. We discuss the potential genesis of this correlation, suggesting that if the initial conditions of Upper Sco resembled the $1-2$~Myr Orion Nebula Cluster, then it was likely imparted at birth as part of the star formation process ($\S$\ref{sec:discussion}).

Next, using a forward-modeling technique, we found that several classes of angular momentum evolution models generically predict the evolution of rotation from Upper Sco at 10~Myr to the Pleiades at 125~Myr, for stars in the mass range 0.1--0.6\msun. The accuracy of our predictions are very weakly dependent on the treatment of internal AM transport, the details of the magnetized wind model, and the method used for determining stellar masses. It remains unclear if the spread of rotation rates about the mean trend in the two clusters correspond, but this situation may be rectified by superior reddening corrections, binary statistics, and membership of the younger cluster. {\it Gaia}~DR2 will undoubtedly mitigate these issues.

Finally, we found that in order to accurately predict the rotation pattern among the M~dwarfs in the Pleiades, a mass-rotation correlation of the magnitude and sign found in Upper Sco must have been imprinted by 10~Myr. We tested this by adopting different initial conditions for our forward models, and found very poor agreement with the empirical Pleiades pattern if an Upper-Sco-like mass-rotation correlation is not present at 10~Myr. This suggests that the initial rotation conditions for the Pleiades were very similar to what we find in the 10~Myr cluster, and thus Upper Sco and the Pleiades form a rotational evolutionary sequence.

\section{Acknowledgements}

We thank Lynne Hillenbrand for comments on the manuscript, Keivan Stassun for helpful advice on the determinations of stellar masses, Greg Herczeg and Fang Qiliang for providing their Upper Scorpius extinction values, and the anonymous referee for valuable input and suggestions. G.S. acknowledges the support of the Vanderbilt Office of the Provost through the Vanderbilt Initiative in Data-intensive Astrophysics (VIDA) fellowship.

\software{YREC \citep{demarque2008}, rotevol \citep{vansaders2013}}

\bibliographystyle{aasjournals}
\bibliography{biblio}

\end{document}